\documentclass[twocolumn,showpacs,preprintnumbers,amsmath,amssymb,APSl,prd,superscriptaddress]{revtex4-1}

\usepackage{dcolumn}
\usepackage{bm}
\usepackage{ifpdf}
\usepackage{hyperref}
\usepackage{dcolumn}
\usepackage{bm}
\usepackage[spanish,english]{babel}
\usepackage{amsfonts}
\usepackage{amssymb}
\usepackage{graphicx}
\usepackage[latin1]{inputenc}

\begin{document}

\title{Reissner-Nordstr\"om Black Holes in Extended Palatini Theories}

\author{Gonzalo J. Olmo} \email{ gonzalo.olmo@csic.es}
\affiliation{Departamento de F\'{i}sica Te\'{o}rica and IFIC, Centro Mixto Universidad de
Valencia - CSIC. Universidad de Valencia, Burjassot-46100, Valencia, Spain}
\author{D. Rubiera-Garcia} \email{rubieradiego@gmail.com}
\affiliation{Departamento de F\'{i}sica, Universidad de Oviedo, Avenida Calvo Sotelo 18, 33007, Oviedo, Asturias, Spain}


\date{\today}

\begin{abstract}

We study static, spherically symmetric solutions with an electric field in an extension of general relativity (GR) containing a Ricci-squared term and formulated in the Palatini formalism. We find that all the solutions present a central core whose area is proportional to the Planck area times the number of charges. Far from the core, curvature invariants quickly tend to those of the usual  Reissner-Nordstr\"om solution, though the structure of horizons may be different. In fact, besides the structures found in the Reissner-Nordstr\"om solution of GR,  we find black hole solutions with just one nondegenerate horizon (Schwarzschild-like), and nonsingular black holes and naked cores.  The charge-to-mass ratio of the nonsingular solutions implies that the core matter density is independent of the specific amounts of charge and mass and of order the Planck density. We discuss the physical implications of these results for astrophysical and microscopic black holes, construct the Penrose diagrams of some illustrative cases, and show that the maximal analytical extension of the nonsingular solutions implies a bounce of the radial coordinate.

\end{abstract}

\maketitle

\section{Introduction}

According to General Relativity (GR), the fate of any sufficiently massive star is to end up its lifetime forming a black hole, a spacetime region that contains a zero-volume singularity of infinite density cloaked by an event horizon. Singularity and uniqueness theorems, together with the cosmic censorship conjecture \cite{singularity}, tell us that black holes are described by solely three parameters: mass, charge and angular momentum, a result known as the no-hair theorem that yields the Kerr-Newman family \cite{MTW}. For any black hole, the event horizon acts as a sort of no-way-out layer that separates the physics outside  the horizon, which is the one that we can directly explore and find it in excellent agreement with the GR predictions \cite{Will05}, from the physics of its interior, where much less is known. When quantum phenomena come into play, the mere existence of an event horizon induces the emission of thermal particles via Hawking radiation \cite{Hawking74}. Quantum effects may also affect the details of gravitational collapse, as recently studied in \cite{Greenwood08}, and are likely to play a very important role when the spacetime curvature reaches the Planck scale. In fact, it is generally accepted that classical black hole singularities should be removed by quantum gravitational effects. However, our current understanding of quantum gravity is not mature enough to provide a clear and detailed description of how this could occur. It seems thus justified to explore phenomenological approaches to this problem hoping to gain some new insights that help us to better understand how black hole structure could be modified by quantum gravity effects.

As the singularity theorems only state some precise physical conditions under which the appearance of singularities is unavoidable, several approaches have been carried out to find conditions that avoid or ameliorate the formation of black hole singularities.  A well known example is Bardeen's black hole \cite{Bardeen68}, in which \emph{exotic} matter sources get rid of the singularity while keeping the horizons and the asymptotically flat character. This singularity avoidance is realized through the formation of a central matter (de Sitter) core, such that the corresponding spacetime is interpreted as the gravitational field of a nonlinear magnetic monopole and can be derived from a nonlinear electrodynamics model \cite{ABG98}. That approach became a prototype  for most developments on regular black holes within classical GR. In this sense, there has been much activity aimed at finding alternative matter sources for the interior region, such as introducing nonlinearities as in the case of nonlinear theories of  electrodynamics \cite{regular-NEDs}, implementing a de Sitter core that matches the exterior field in some ``junction" region  (see e.g. \cite{Lemos11, Ansoldi08} and references therein), or using new ideas inspired by noncommutative geometry \cite{Nicolini06}, in such a way that singularities are removed. Also other regular magnetically charged solutions within GR have been found \cite{Bronnikov01}.

A different approach comes from the idea that extensions of GR with high-curvature corrections could be able to capture some essential features useful or required to find an effective description of the quantum gravity dynamics. This point is supported by the study of quantum fields in curved spacetimes \cite{P&T_B&D} and by approaches to quantum gravity  based on string theory \cite{Ortin, strings}. In this sense, gravity theories containing higher-order curvature invariants naturally appear as modifications of GR in these quantum gravity approaches \cite{P&T_B&D, Ortin, strings}. Such theories generically lead to higher-order partial derivative equations, which is a manifestation of the fact that new high-energy degrees of freedom are being taken into account. The resulting solutions are thus expected to depend on a larger number of parameters (integration constants), which should provide more freedom and/or new mechanisms to avoid the singularities. However, nonsingular black holes of this kind are rather scarce (see e.g. \cite{Schmidt85, Berej06}) and usually require the addition of exotic sources of matter again. This suggests that the addition of new degrees of freedom in the problem is unable by itself to resolve the problem of singularities. Moreover, these theories suffer from ghosts and other perturbative instabilities \cite{Zumino86}. These problems, however, can be avoided if the curvature invariants appear in appropriate combinations, because then the equations of motion may remain second-order like in GR. These are known as Lovelock gravities \cite{Lovelock} (see, for instance, \cite{Zanelli} for a pedagogical introduction).  For example, the simplest extension of Einstein gravity via higher-curvature terms in this context corresponds to Gauss-Bonnet gravity and picks up three new terms, $R_{\mu\nu\rho\sigma}R^{\mu\nu\rho\sigma}-4R_{\mu\nu}R^{\mu\nu}+R^2$, which supplement the Einstein-Hilbert Lagrangian. It should be noted that in $3+1$ dimensions these new terms are topological invariants not contributing to the equations of motion, which means that Lovelock gravities only provide modified dynamics in the context of higher dimensions. Exact static spherically symmetric solutions to Gauss-Bonnet theory in vacuum \cite{Boulware85} and with electrostatic fields \cite{Wiltshire88} are known, but they still contain singularities or are ill defined.

An alternative strategy to obtain modified gravitational dynamics beyond GR is to assume that the metric and affine structures of the theory are independent \cite{Zanelli}. When the connection is not constrained {\it a priori} to be given by the Christoffel symbols of the metric,  one finds that even $f(R)$ extensions yield second-order field equations, which contrasts with the usual (metric) formulation of those theories \cite{Met-Reviews} and the general belief that only Lovelock gravities have second-order equations. This approach, known as Palatini formalism, has been recently used to obtain a covariant action \cite{O&S2009} for the effective Hamiltonian dynamics of loop quantum cosmology \cite{LQC}, an approach to quantum cosmology based on the nonperturbative quantization techniques of loop quantum gravity \cite{Loop}. In the Palatini approach \cite{Pal-Review}, metric and connection are regarded as independent entities and the field equations are obtained by independent variation of the action with respect to both of them. Though this does not affect the dynamics of GR \cite{MTW},
it does have important consequences for extensions of it \cite{Pal-Review}. In general, one finds that in Palatini theories the matter plays an active role in the construction of the independent connection, which ends up producing modified dynamics. When there is no matter, the field equations boil down to those of GR with an effective cosmological constant, which depends on the form of the particular Lagrangian chosen. This property has made these theories very attractive for cosmological applications.

The unusual role played by the matter in the construction of the geometry in Palatini theories together with the second-order character of their field equations makes them specially interesting to explore the effects of new gravitational physics on the structure of black holes. In this sense, in Ref.\cite{OR2011a} we considered Palatini $f(R)$ modifications of GR in interaction with modified matter sources, through nonlinear electrodynamics (NEDs).  In Palatini $f(R)$ theories, the modified dynamics is due to a number of new terms on the right-hand side of the equations that depend on the  trace $T$ of the energy-momentum tensor of the matter. Unlike Maxwell's electrodynamics, the stress-energy tensor of NEDs possesses, in general, a nonvanishing trace, which makes them specially suitable to excite the Palatini modified dynamics in electrovacuum scenarios.  In the particular case of Born-Infeld NED coupled to the gravity theory $f(R)=R\pm R^2/R_P$, where $R_P$ is the Planck curvature, we found that the degree of divergence of the Kretschmann scalar near the singularity can be much weaker than in GR.

In this work we go beyond Ref.\cite{OR2011a} and explore how black hole structure is affected by new physics at the Planck scale by considering a Palatini theory of the form $f(R,Q)=R+aR^2/R_P+b Q/R_P$, where $Q\equiv R_{\mu\nu}R^{\mu\nu}$, $R_{\mu\nu}$ is the (symmetric) Ricci tensor, and $a$ and $b$ are constants. Terms of this kind have been considered in the metric approach in an attempt to find ghost and singularity free theories of gravity \cite{Biswas12}, and also in the study of black holes coupled to NEDs \cite{Berej06}. In the Palatini framework, $f(R,Q)$ theories yield second-order field equations that exactly boil down to the usual Einstein-de Sitter equations in vacuum (see details in \cite{Olmo2011QG} and below), which guarantees the absence of  ghosts and other dynamical instabilities. The presence of a Ricci-squared term in the action is very important because it leads to modified dynamics even for traceless matter sources, such as radiation and the usual Maxwell electromagnetic field, which contrasts with $f(R)$ theories. As a result, this quadratic $f(R,Q)$ model provides the simplest extension beyond GR of the usual (nonrotating) Reissner-Nordstr\"om black hole. The quadratic $f(R,Q)$ Palatini model proposed here has already been studied in cosmological scenarios, where it was found that the big bang singularity is replaced by a cosmic bounce in isotropic and anisotropic (Bianchi I) universes filled with standard sources of matter and radiation \cite{Barragan2010}.

In Refs.\cite{OR2011b,OR2012b} we reported on several key aspects of nonsingular black hole  solutions found in this model. In this paper we shall go further on this subject extending those results and providing a comprehensive account of all the derivations and technical details. The paper is organized as follows. In section \ref{section:II} we recall some basic aspects of Palatini $f(R,Q)$ theories and write the associated field equations. The particular actions for matter (Maxwell) and gravity sectors of our theory are introduced in section \ref{section:III}, where we construct all the relevant geometric objects. In section \ref{section:IV} we comment on the choice of ansatz for the line element, and proceed to solve the metric field equations. Section \ref{section:V} is devoted to the analysis of the interior region and to the characterization of the event horizons. In particular we construct the Penrose diagrams and their maximal analytical extensions associated to the different black holes and {\it naked cores} found. The physical aspects of these solutions, including singular and nonsingular black holes,  and microscopic {\it naked cores}, are discussed in section \ref{section:VI}. We conclude in section \ref{section:VII} with a summary and some future perspectives.

\section{Action and field equations} \label{section:II}

We define Palatini $f(R,Q)$ theories as follows
\begin{equation}\label{eq:action}
S[g,\Gamma,\psi_m]=\frac{1}{2\kappa^2}\int d^4x \sqrt{-g}f(R,Q) +S_m[g,\psi_m],
\end{equation}
where $\kappa^2\equiv 8\pi G$,  $S_m[g,\psi_m]$ represents the matter action, $g_{\alpha\beta}$ is the space-time metric, $R=g^{\mu\nu}R_{\mu\nu}$, $Q=g^{\mu\alpha}g^{\nu\beta}R_{\mu\nu}R_{\alpha\beta}$, $R_{\mu\nu}={R^\rho}_{\mu\rho\nu}$, and
\begin{equation}\label{eq:Riemann}
{R^\alpha}_{\beta\mu\nu}=\partial_{\mu}
\Gamma^{\alpha}_{\nu\beta}-\partial_{\nu}
\Gamma^{\alpha}_{\mu\beta}+\Gamma^{\alpha}_{\mu\lambda}\Gamma^{\lambda}_{\nu\beta}-\Gamma^{\alpha}_{\nu\lambda}\Gamma^{\lambda}_{\mu\beta} \ .
\end{equation}

Variation of (\ref{eq:action}) with respect to metric and connection leads to the following equations \cite{OSAT09}
\begin{eqnarray}
f_R R_{\mu\nu}-\frac{f}{2}g_{\mu\nu}+2f_QR_{\mu\alpha}{R^\alpha}_\nu &=& \kappa^2 T_{\mu\nu}\label{eq:met-varX}\\
\nabla_{\beta}\left[\sqrt{-g}\left(f_R g^{\mu\nu}+2f_Q R^{\mu\nu}\right)\right]&=&0  \ ,
 \label{eq:con-varX}
\end{eqnarray}
were we have used the short-hand notation $f_X\equiv \partial_X f$. In the above derivation we have assumed a symmetric Ricci tensor, $R_{\mu\nu}=R_{\nu\mu}$, and vanishing torsion. The condition on the Ricci tensor,
$R_{[\mu\nu]}=0$, forces the connection components $\Gamma^\sigma_{\sigma\nu}$ to be the gradient of a scalar function, $\Gamma^\sigma_{\sigma\nu}=\partial_\nu \phi$. In the usual formulation of GR, where the connection is given by the Christoffel symbols of the metric, one finds that $\Gamma^\alpha_{\alpha \mu}=\partial_\mu \ln \sqrt{-g}$. In our theory (\ref{eq:action}), the condition $R_{[\mu\nu]}=0$ is equivalent to assuming that $\Gamma^{\alpha}_{\beta\gamma}$ can be solved as the Levi-Civita connection of an auxiliary metric $h_{\mu\nu}$, which leads to $\Gamma^\alpha_{\alpha \mu}=\partial_\mu \ln \sqrt{-h}$. The explicit relation between $h_{\mu\nu}$ and $g_{\mu\nu}$ follows from the field equations and will be discussed later. A reason to set $R_{[\mu\nu]}=0$ is that then the field equations of (\ref{eq:action}) in vacuum boil down exactly to those of GR (with possibly a cosmological constant, depending on the function $f(R,Q)$ chosen). This guarantees that there are no new propagating degrees of freedom and, therefore, the resulting theory is not affected by ghosts or other dynamical instabilities. In regions containing sources, the equations of (\ref{eq:action})  differ from those of GR due to the presence of new matter/energy-dependent terms induced by the mismatch between $h_{\mu\nu}$ and $g_{\mu\nu}$, which leads to modified gravitational dynamics without introducing new dynamical degrees of freedom or higher-order derivatives of the metric. If the condition $R_{[\mu\nu]}=0$ is relaxed, then $\Gamma^\alpha_{\alpha \mu}$ must also have a purely vectorial component, $\Gamma^\alpha_{\alpha \mu}=\partial_\mu\phi(x)+B_\mu$,  which adds new dynamical degrees of freedom to the theory.  In that case, the dynamics of (\ref{eq:action}) differs from that of GR even in the absence of matter/energy sources (see \cite{Ricci_subtlety} for related results in this direction).

We now focus on working out a solution for (\ref{eq:con-varX}). At first sight,  since $R$ and $R_{\mu\nu}$ are functions of the connection and its first derivatives, (\ref{eq:con-varX}) can be seen as a nonlinear, second-order partial differential equation for the unknown connection. However, there exist algebraic relations between $R$, $R_{\mu\nu}$ and the energy-momentum tensor of the matter that make the problem easier to handle.
To proceed, we first define the matrix $\hat{P}$, whose components are ${P_\mu}^\nu\equiv R_{\mu\alpha}g^{\alpha\nu}$, which allows us to express (\ref{eq:met-varX}) as
\begin{equation}
f_R {P_\mu}^\nu-\frac{f}{2}{\delta_\mu}^\nu+2f_Q{P_\mu}^\alpha {P_\alpha}^\nu= \kappa^2 {T_\mu}^\nu\label{eq:met-varRQ1} \ .
\end{equation}
In matrix notation, this equation reads
\begin{equation}
2f_Q\hat{P}^2+f_R \hat{P}-\frac{f}{2}\hat{I} = \kappa^2 \hat{T} \label{eq:met-varRQ2} \ ,
\end{equation}
where $\hat{T}$ is the matrix representation of ${T_\mu}^\nu$. Note that $R$ and $Q$ are the trace of $\hat{P}$ and  $\hat{P}^2$, respectively. The solution of (\ref{eq:met-varRQ2}) implies that  $\hat{P}$ can be expressed as a function of the components of the energy-momentum tensor, i.e.,  $\hat{P}=\hat{P}(\hat{T})$. Assuming that for a given $f(R,Q)$ theory such a solution exists, Eq.(\ref{eq:con-varX}) can now be seen as an algebraic equation for the connection in which, besides the metric $g_{\mu\nu}$, there is an explicit dependence on the energy-momentum tensor of the matter. To solve it,
we look for a metric $\hat{h}$ such that (\ref{eq:con-varX}) becomes $\nabla_{\beta}\left[\sqrt{-h} h^{\mu\nu}\right]=0$. This guarantees that the independent connection can be expressed as the Levi-Civita connection of $\hat{h}$. Using matrix notation, we have
\begin{equation}
\sqrt{-h}\hat{h}^{-1}=\sqrt{-g}\hat{g}^{-1}\hat\Sigma \ ,
\end{equation}
where we have defined $\hat\Sigma=\left(f_R\hat{I}+2f_Q\hat{P}\right)$.
Computing the determinant of this expression, we find $h=g \det\hat\Sigma$. With this result, we have
\begin{equation} \label{eq:h-g}
\hat{h}^{-1}=\frac{\hat{g}^{-1}\hat\Sigma}{\sqrt{\det\hat\Sigma}} \ , \
\hat{h}=\left(\sqrt{\det\hat\Sigma}\right)\hat\Sigma^{-1}\hat{g} \ .
\end{equation}
This shows that the connection of $f(R,Q)$ theories can be explicitly solved in terms of the physical metric $g_{\mu\nu}$ and the matter sources.

With the above results, the metric field equations can be rewritten in a more compact and transparent form. Expressing  (\ref{eq:met-varRQ2}) as
\begin{equation}
\hat{P}\hat{\Sigma} =\frac{f}{2}\hat{I} +\kappa^2 \hat{T} \label{eq:met-varRQ3} \ ,
\end{equation}
and using (\ref{eq:h-g}), we can rewrite ${P_\mu}^\alpha {\Sigma_\alpha}^\nu$ as $ R_{\mu\alpha}h^{\alpha\nu} \sqrt{\det \hat\Sigma}$. This allows to express   (\ref{eq:met-varRQ3}) as
\begin{equation}
{R_\mu}^\nu (h) =\frac{1}{\sqrt{\det \hat\Sigma}}\left(\frac{f}{2}{\delta_\mu}^\nu +\kappa^2 {T_\mu}^\nu\right) \label{eq:met-varRQ4} \ ,
\end{equation}
where  ${T_\mu}^\nu=T_{\mu\alpha}g^{\alpha \nu}$.

\section{Electrically charged $f(R,Q)$ black holes} \label{section:III}

\subsection{Matter Lagrangian}

The Schwarzschild black hole is the most general spherically symmetric, nonrotating vacuum solution of GR
and also of (\ref{eq:met-varX}). However, that solution assumes that all the matter is concentrated on a point
of infinite density, which is not consistent with the dynamics of (\ref{eq:met-varX}). In fact, if one considers
the collapsing object as described by a perfect fluid that behaves as radiation during the last stages of the collapse, explicit computation of the scalar $Q=R_{\mu\nu}R^{\mu\nu}$ shows that the energy density $\rho$ is bounded from above by $\rho_{max}=\rho_P/32$, where $\rho_P\equiv 3 c^5/4\pi \hbar G^2 \approx 10^{94}$ g/cm$^3$ is Planck's density \cite{Barragan2010,OSAT09}. Therefore, one should study the complicated process of collapse of a spherical nonrotating object to determine how the Schwarzschild metric is modified in our theory. For this reason we study instead vacuum space-times with an electric field, which possess a nonzero stress-energy tensor able to excite the Palatini dynamics even in static settings. The resulting solutions should therefore be seen as Planck-scale modifications of the usual Reissner-Nordstr\"om solution. Let us thus consider as the matter source in action (\ref{eq:action}) the Maxwell lagrangian

\begin{equation}
S_m[g,\psi_m] = -\frac{1}{2\kappa^2}\int d^4x \sqrt{-g}F_{\alpha\beta}F^{\alpha\beta}
\end{equation}
whose associated stress-energy tensor is written as

\begin{equation} \label{eq:em}
{T_\mu}^\nu=-\frac{1}{4\pi}\left[{F_\mu}^\alpha {F_\alpha}^\nu-\frac{{F_\alpha}^\beta{F_\beta}^\alpha}{4}\delta_\mu^\nu\right] \ ,
\end{equation}
where $F_{\mu\nu}=\partial_\mu A_\nu-\partial_\nu A_\mu$ is the field strength tensor of the vector potential $A_{\mu}$. For a purely electrostatic configuration and taking a spherically symmetry line element of the form $ds^2=g_{tt}dt^2+g_{rr}dr^2+r^2d\Omega^2$, with $d\Omega^2=d\theta^2 + \sin^2 \theta d\varphi^2$, one finds that the only nonvanishing component is $F^{tr}$. It is then easy to see that the (sourceless) field equations $\nabla_\mu F^{\mu\nu}=0$ lead to
\begin{equation}
F^{tr}=\frac{q}{r^2}\frac{1}{\sqrt{-g_{tt}g_{rr}}} \ ,
\end{equation}
where $q$ is an integration constant that represents the charge generating the electric field. With this result, the product ${F_\mu}^\alpha {F_\alpha}^\nu$ in (\ref{eq:em}) becomes
\begin{equation}
{F_\mu}^\alpha {F_\alpha}^\nu=\begin{pmatrix}
 -g_{tt}g_{rr}(F^{tr})^2\hat{I} & \hat{0}  \\
\hat{0} & \hat{0}
\end{pmatrix}=\begin{pmatrix}
 \frac{q^2}{r^4}\hat{I} & \hat{0}  \\
\hat{0} & \hat{0}
\end{pmatrix} \ ,
\end{equation}
where $\hat{I}$ and $\hat{0}$ represent the $2\times 2$ identity and zero matrices, respectively. Using this, we find that
\begin{equation}\label{eq:Tmn-EM}
{T_\mu}^\nu=\frac{q^2}{8\pi r^4}\begin{pmatrix}
 -\hat{I} & \hat{0}  \\
\hat{0} & \hat{I}
\end{pmatrix}.
\end{equation}

In order to write the field equations in the form (\ref{eq:met-varRQ4}), we first need to find the explicit form of $\hat{P}$ for this matter source, which will allow us to construct $\hat{\Sigma}$ and compute its determinant. To do this, we use (\ref{eq:Tmn-EM})  and write (\ref{eq:met-varRQ2}) as
\begin{equation}
2f_Q\left(\hat{P}+\frac{f_R}{4f_Q}\hat{I}\right)^2=
\begin{pmatrix}
\lambda_-^2\hat{I} & \hat{0}  \\
\hat{0} & \lambda_+^2\hat{I}
\end{pmatrix}  \ ,
\end{equation}
where $\lambda_{\pm}^2=\left(f+\frac{f_R^2}{4f_Q}\pm \frac{\tilde{\kappa}^2q^2}{r^4}\right)/2$ and we have defined $\tilde{\kappa}^2=\kappa^2/4\pi=2G$.
It is easy to see that there are $16$ square roots that satisfy this equation, namely,
\begin{equation}
\sqrt{2f_Q}\left(\hat{P}+\frac{f_R}{4f_Q}\hat{I}\right)=
\begin{pmatrix}
s_1\lambda_- & 0 & 0 & 0  \\
0 & s_2 \lambda_- & 0 & 0\\
0 & 0 & s_3\lambda_+ & 0 \\
0 & 0 & 0 & s_4\lambda_+
\end{pmatrix}  \ ,
\end{equation}
where $s_i=\pm 1$. Agreement with GR in the low curvature regime (where $f_R\to 1$ and $f_Q\to 0$)  requires $s_i=+1$. For this reason, we simplify the notation and take
\begin{equation}\label{eq:M_ab}
\sqrt{2f_Q}\left(\hat{P}+\frac{f_R}{4f_Q}\hat{I}\right)=
\begin{pmatrix}
\lambda_- \hat{I}& \hat{0} \\
\hat{0} & \lambda_+ \hat{I}
\end{pmatrix}  \ .
\end{equation}
From this it follows that the matrix $\hat{\Sigma}$ is given by
\begin{equation}
\hat{\Sigma}=\frac{f_R}{2}\hat{I}+\sqrt{2f_Q}
\begin{pmatrix}
\lambda_- \hat{I}& \hat{0} \\
\hat{0} & \lambda_+ \hat{I}
\end{pmatrix}=\begin{pmatrix}
\sigma_- \hat{I}& \hat{0} \\
\hat{0} & \sigma_+\hat{I}
\end{pmatrix} \ ,
\end{equation}
where $\sigma_\pm=\left(\frac{f_R}{2}+\sqrt{2f_Q}\lambda_\pm\right)$.
From this expression it is easy to see that $\sqrt{\det{\hat\Sigma}}=\sigma_+\sigma_-$ and, therefore, the field equations (\ref{eq:met-varRQ4}) become
\begin{equation}
{R_\mu}^\nu(h)=\frac{1}{2\sigma_+\sigma_-}\begin{pmatrix}
\left(f-\frac{\tilde{\kappa}^2q^2}{r^4}\right) \hat{I}& \hat{0} \\
\hat{0} & \left(f+\frac{\tilde{\kappa}^2q^2}{r^4}\right)\hat{I}
\end{pmatrix}  \label{eq:Ricci-h3} \ .
\end{equation}

\subsection{Gravity Lagrangian}

To work out the explicit form of the metric we must specify an $f(R,Q)$ model. It is very useful to consider the family $f(R,Q)=\tilde{f}(R)+ l_P^2 Q$, where $l_P=\sqrt{\hbar G/c^3}\sim 10^{-35}m$ is Planck's length, because tracing (\ref{eq:met-varX}) with the metric $g^{\mu\nu}$ leads to the well known $f(R)$ relation $R \tilde{f}_R-2\tilde{f}=\kappa^2 T$, which implies that $R=R(T)$. Since for Maxwell theory $T=0$, it follows that the $\tilde{f}(R)$ part of the $f(R,Q)$ theory does not play a very relevant role in the dynamics. We will just assume that the $\tilde{f}(R)$ part is close to GR, $\tilde{f}(R)=R+a_2 R^2+a_3R^3+\ldots$, and that $R(T=0)=0$ for simplicity (and for consistency with the choice $s_i=+1$ above). This is true, in particular, for the model
\begin{equation}
f(R,Q)=R+l_P^2(a R^2+Q) \ ,
\end{equation}
whose cosmological dynamics has been carefully studied in the literature \cite{Barragan2010} and that we set as the model to be discussed from now on. For this model we have that when ${T_{\mu}}^{\nu}$ is given by (\ref{eq:Tmn-EM}) then $R=0$, $f_R=1$, and $f(R,Q)=l_P^2 Q$. Using this in (\ref{eq:M_ab}) and taking the trace we find
\begin{equation}
1=\sqrt{\frac{1}{4}+l_P^4 Q+\frac{\tilde{\kappa}^2q^2l_P^2}{ r^4}}+\sqrt{\frac{1}{4}+l_P^4{Q}-\frac{\tilde{\kappa}^2q^2l_P^2}{r^4}} \ ,
\end{equation}
from which we obtain
\begin{equation}\label{eq:Q}
Q=\frac{\tilde{\kappa}^4 q^4}{r^8} \ ,
\end{equation}
which coincides with the expression of GR. From this result it follows that $\lambda_{\pm}= \frac{1}{l_P\sqrt{2}}\left(\frac{1}{2}\pm\frac{\tilde{\kappa}^2q^2l_P^2}{r^4}\right)$ and $\sigma_{\pm}=1\pm \frac{\tilde{\kappa}^2q^2l_P^2}{r^4}$.
Noting also that $f\pm\frac{\tilde{\kappa}^2q^2}{r^4}=\pm\frac{\tilde{\kappa}^2q^2}{r^4} \sigma_\pm$, (\ref{eq:Ricci-h3}) becomes
\begin{equation}
{R_\mu}^\nu(h)=\frac{\tilde{\kappa}^2q^2}{2r^4}\begin{pmatrix}
-\frac{1}{\sigma_+} \hat{I}& \hat{0} \\
\hat{0} & \frac{1}{\sigma_-} \hat{I}
\end{pmatrix}  \label{eq:Ricci-h4} \ .
\end{equation}
which fully determines the dynamics of our $f(R,Q)$ theory coupled to Maxwell electrodynamics. These equations exactly recover GR in the limit $l_P\to 0$.

\section{Solving the field equations} \label{section:IV}

In order to solve for the metric $g_{\mu\nu}$, it is more convenient to solve first for $h_{\mu\nu}$ using  (\ref{eq:Ricci-h4}) and then transform  back to $g_{\mu\nu}$ using the relation $g_{\mu\nu}={\Sigma_\mu}^\alpha h_{\alpha\nu}/\sqrt{\det\hat \Sigma}$ that follows from (\ref{eq:h-g}). To do this, it is convenient to define a line element associated to the metric $h_{\mu\nu}$ using a set of Schwarzschild-like coordinates as follows
\begin{equation}
d\tilde{s}^2=h_{tt}dt^2+h_{\tilde{r}\tilde{r}}d\tilde{r}^2+\tilde{r}^2d\Omega^2 \label{eq:ds2h} \ .
\end{equation}
This line element is formally identical to that corresponding to the physical metric $g_{\mu\nu}$,
\begin{equation}
d{s}^2=g_{tt}dt^2+g_{{r}{r}}d{r}^2+{r}^2d\Omega^2 \label{eq:ds2g} \ ,
\end{equation}
but their relation is not trivial due to the nonconformal relation between the two metrics and the different choice of radial coordinate $r\neq \tilde{r}$. This can be seen from the relation $g_{\mu\nu}={\Sigma_\mu}^\alpha h_{\alpha\nu}/\sqrt{\det \hat\Sigma}$, which leads to
\begin{equation}\label{eq:g-vs-h}
g_{\mu\nu}=\begin{pmatrix}
g_{tt} & 0 & 0 & 0 \\
0 & g_{rr} & 0 & 0 \\
0 & 0 & r^2 & 0 \\
0 & 0 & 0 & r^2\sin^2 \theta
\end{pmatrix} = \begin{pmatrix}
\frac{h_{tt}}{\sigma_+} & 0 & 0 & 0 \\
0 & \frac{h_{{r}{r}}}{\sigma_+} & 0 & 0 \\
0 & 0 & \frac{h_{\theta\theta}}{\sigma_-} & 0 \\
0 & 0 & 0 & \frac{h_{\phi\phi}}{\sigma_-}
\end{pmatrix} \ .
\end{equation}
From the line element (\ref{eq:ds2h}) it is easy to see that $g_{\theta\theta}=h_{\theta\theta}/\sigma_-$ implies that $\tilde{r}^2=r^2 \sigma_-= r^2-\frac{\tilde{\kappa}^2q^2l_P^2}{r^2}$. It is also easy to see that $g_{rr}=h_{{r}{r}}/\sigma_+=(h_{\tilde{r}\tilde{r}}/\sigma_+)(d\tilde{r}/dr)^2$.
Since the time coordinate is the same in the two line elements, for the $g_{tt}$ component we just have $g_{tt}=h_{tt}/\sigma_+$.

It should be noted that, in general, the line elements can be written without using $r$ and $\tilde{r}$ as coordinates \cite{Stephani2003}. In that case, we would have $d\tilde{s}^2=h_{ab}(x^0,x^1)dx^a dx^b+\tilde{r}^2(x^0,x^1)d\Omega^2$ and  $d{s}^2=g_{ab}(x^0,x^1)dx^a dx^b+{r}^2(x^0,x^1)d\Omega^2$, with the relations $g_{ab}(x^0,x^1)=h_{ab}(x^0,x^1)/\sigma_+$ and $\tilde{r}^2=r^2 \sigma_-$, being the latter clearly independent of the choice of $(x^0,x^1)$.  At this point, it is important to note that the radial function $\tilde{r}^2$ vanishes when $r^4= \tilde{\kappa}^2q^2l_P^2$. This means that the $2-$spheres of the $h-$geometry (the geometry associated to the independent connection)  can only be put into correspondence with the $2-$spheres of the $g-$geometry up to $r/(\tilde{\kappa}|q|l_P)^{1/2}\ge 1$. Since the $h-$geometry deviates from the $g-$geometry by the effects of a matter-induced deformation (represented by the matrix ${\Sigma_\mu}^\nu$), the impossibility of mapping any portion of the $h-$geometry into the $r/(\tilde{\kappa}|q|l_P)^{1/2}<1$ sector of the $g-$ geometry suggests that the matter (the electromagnetic field in this case) cannot penetrate in that region of the physical spacetime. The description and analysis of the properties of the hypersurface $r/(\tilde{\kappa}|q|l_P)^{1/2}=1$ will be one of the main goals of this paper.

Working with the Schwarzschild (canonical) coordinates of above and
introducing the ansatz $h_{tt}=-A(\tilde{r}) e^{2\psi(\tilde{r})}$ and $h_{\tilde{r}\tilde{r}}=1/A(\tilde{r})$, the components of the tensor ${R_\mu}^\nu(h)$ become
\begin{eqnarray}\label{eq:Rtt}
{R_t}^t&=&-\frac{1}{2h_{\tilde{r}\tilde{r}}}\Bigg[\frac{A_{\tilde{r}\tilde{r}}}{A}-\left(\frac{A_{\tilde{r}}}{A}\right)^2+
2\psi_{\tilde{r}\tilde{r}}+ \nonumber \\ &+& \left(\frac{A_{\tilde{r}}}{A}+ 2\psi_{\tilde{r}}\right)\left(\frac{A_{\tilde{r}}}{A}+
\psi_{\tilde{r}}+\frac{2}{\tilde{r}}\right)\Bigg] \\
{R_{\tilde{r}}}^{\tilde{r}}&=&-\frac{1}{2h_{\tilde{r}\tilde{r}}}\Bigg[\frac{A_{\tilde{r}\tilde{r}}}{A}-
\left(\frac{A_{\tilde{r}}}{A}\right)^2+2\psi_{\tilde{r}\tilde{r}}+ \nonumber \\ &+& \left(\frac{A_{\tilde{r}}}{A}+
2\psi_{\tilde{r}}\right)\left(\frac{A_{\tilde{r}}}{A}+\psi_{\tilde{r}}\right)+
\frac{2}{\tilde{r}}\frac{A_{\tilde{r}}}{A}\Bigg] \label{eq:Rrr}\\
{R_\theta}^\theta&=&\frac{1}{\tilde{r}^2}\left[1-A(1+\tilde{r}\psi_{\tilde{r}})-\tilde{r}A_{\tilde{r}}\right] \ . \label{eq:Rzz}
\end{eqnarray}
From (\ref{eq:Ricci-h4}) it is easy to see that the combination ${R_t}^t-{R_{\tilde{r}}}^{\tilde{r}}=0$. Using (\ref{eq:Rtt}) and (\ref{eq:Rrr}), this combination implies that $\psi_{\tilde{r}}=0 \rightarrow \psi= constant$, like in GR and $f(R)$ theories. As usual, this constant can be eliminated by a redefinition of the time coordinate, leaving a single function ($A(\tilde{r})$) to be determined. From (\ref{eq:Rzz}) we find that $A(\tilde{r})$ satisfies the following equation
\begin{equation}\label{eq:A(r)}
\frac{1}{\tilde{r}^2}\left[1-A-\tilde{r}A_{\tilde{r}}\right]=\frac{\tilde{\kappa}^2q^2}{2r^2}\frac{1}{\left(r^2-\frac{\tilde{\kappa}^2q^2l_P^2}{r^2}\right)}
\end{equation}
Using the ansatz $A(\tilde{r})=1-2 M(\tilde{r})/\tilde{r}$ and the relation $\tilde{r}^2= r^2-\frac{\tilde{\kappa}^2q^2l_P^2}{r^2}$, (\ref{eq:A(r)}) turns into
\begin{equation}\label{eq:M(r)}
M_{\tilde{r}}=\frac{\tilde{\kappa}^2q^2}{4r^2} \ .
\end{equation}
Taking into account that $d\tilde{r}/dr=\sigma_+/\sigma_-^{1/2}$, the above expression becomes
\begin{equation}\label{eq:M(r)}
M_r=\frac{\tilde{\kappa}^2q^2\sigma_+}{4r^2 \sigma_-^{1/2}} \ ,
\end{equation}
which reduces the problem to solving a first-order differential equation in the variable $r$. The integration constant of this equation can be identified with the Schwarzschild mass $M_0\equiv r_S/2$ of the vacuum problem ($q=0$). We can thus write $2M(r)=r_S+\Delta M$ to emphasize that it is the function $\Delta M$ which encodes the  electrostatic contribution to the mass function. In order to obtain $\Delta M$, it is useful to introduce some definitions to work with dimensionless variables. We thus define a length (squared) associated to the charge,  $r_q^2\equiv \tilde{\kappa}^2 q^2$, and a dimensionless radial variable $z\equiv r/ \sqrt{r_q l_P}$. With this notation, the metric $g_{\mu\nu}$ can  be expressed as
\begin{equation}\label{eq:g}
g_{tt}=-\frac{A(z)}{\sigma_+} \ , \ g_{rr}=\frac{\sigma_+}{\sigma_-A(z)}  \ , \ A(z)=1-\frac{\left[1+\delta_1 G(z)\right]}{\delta_2 z \sigma_-^{1/2}} \ .
\end{equation}
where $\sigma_\pm=1\pm1/z^4$, we used the relation $\tilde{r}=r \sigma_{-}^{1/2}$ and defined $\Delta M/r_S=\delta_1 G(z)$, together with
\begin{equation}\label{eq:d1d2}
\delta_1=\frac{1}{2r_S}\sqrt{\frac{r_q^3}{l_P}} \ , \
\delta_2= \frac{\sqrt{r_q l_P} }{r_S} \ .
\end{equation}
The (dimensionless) function $G(z)$ has a purely electrostatic origin and satisfies the following equation
\begin{equation} \label{eq:Gz}
\frac{dG}{dz}=\frac{z^4+1}{z^4\sqrt{z^4-1}} \ .
\end{equation}
We stress that the two scales of the problem, namely, the integration constants $q$ and $M_0$, have been replaced by the dimensionless ratios $\delta_1$ and $\delta_2$ given in (\ref{eq:d1d2}).

\subsection{Finding $G(z)$}

The integration of (\ref{eq:Gz}) to obtain the function $G(z)$ can be carried out straightforwardly using power series expansions in two regions of interest. One is the far limit $z\gg 1$ and the other is $z\sim 1$.

\subsubsection{Far limit $z\gg 1$}
To study this limit, it is useful to express $dG/dz$ as follows
\begin{equation}
G_z=\left(\frac{1}{z^2}+\frac{1}{z^6}\right)\frac{1}{(1-1/z^4)^{1/2}} \ .
\end{equation}
Using the binomial expansion $(1+x)^\alpha=\sum_{k=0}^\infty \begin{pmatrix} \alpha \\ k \end{pmatrix}x^k$, the above expression becomes
\begin{equation}
G_z=\sum_{k=0}^\infty (-1)^k\begin{pmatrix} -1/2 \\ k \end{pmatrix}\left(z^{-2-4k}+z^{-6-4k}\right) \ .
\end{equation}
The integration of this expression is immediate and gives
\begin{equation}\label{eq:far}
G(z)=-\frac{1}{z}\sum_{k=0}^\infty \frac{(-1)^k}{(1+4k)z^{4k}}\begin{pmatrix} -1/2 \\ k \end{pmatrix}
\left(1+\frac{(1+4k)}{(5+4k)z^4}\right) \ .
\end{equation}
From the first terms of this expansion,  $G(z)\approx -1/z-3/10z^5$, one can verify that when $r\gg l_P$ the GR limit (i.e. the standard Reissner-Nordstr\"om solution) is recovered
\begin{eqnarray} \label{eq:gtt-far}
g_{tt}&\approx & -\left(1-\frac{r_S}{r}+\frac{r_q^2}{2r^2}\right) +\frac{r_q^2 l_P^2}{r^4}\\
{g_{rr}}&\approx & \left(1-\frac{r_S}{r}+\frac{r_q^2}{2r^2}-2\frac{r_q^2 l_P^2}{r^4} \right) ^{-1} \ , \label{eq:grr-far}
\end{eqnarray}
where the first-order corrections come from the $\sigma_\pm$ functions rather than from the expansion of $G(z)$. As it follows from (\ref{eq:gtt-far}) and (\ref{eq:grr-far}), as long as $r_S\gg l_P$, the location of the external horizon in these black holes is essentially the same as in GR. How this picture changes for microscopic black holes will be discussed later on.

\subsubsection{Near limit $z\to 1$}

To study this limit, we first consider the change of variable $z^4=1+x$, which leads to
\begin{equation}
G_x=\frac{1}{4}\left[x^{-1/2}\left(1+x\right)^{-3/4}+x^{-1/2}\left(1+x\right)^{-7/4}\right] \ .
\end{equation}
Using again the binomial expansion, we find the following solution
\begin{eqnarray}
G(z)&=&\frac{1}{4}\sum_{k=0}^\infty \frac{(z^4-1)^{k+1/2}}{(k+1/2)}\left[\begin{pmatrix} -3/4 \\ k \end{pmatrix}+\begin{pmatrix} -7/4 \\ k \end{pmatrix}\right] \nonumber \\ &\approx & 2 \sqrt{z-1}-\frac{11}{6} (z-1)^{3/2}+O[z-1]^{5/2}.
\end{eqnarray}
To have agreement with the solution (\ref{eq:far}) found before, we need to add an integration constant $\beta$ on the right hand side of this expansion, which leads to
\begin{equation}\label{eq:near}
G(z)=\beta +\frac{1}{2} \sqrt{z^4-1} \left[f_{\frac{3}{4}}(z)+f_{\frac{7}{4}}(z)\right] \ ,
\end{equation}
where $f_{\lambda}(z)= {_{2}F}_1[\frac{1}{2},\lambda,\frac{3}{2},1-z^4]$ is a hypergeometric function, and $\beta\approx -1.74804$.

The mass function can then be written as
\begin{equation}\label{eq:Near_series}
\frac{M(z)}{M_0}=1+\delta_1\left(\beta +\frac{1}{2} \sqrt{z^4-1} \left[f_{\frac{3}{4}}(z)+f_{\frac{7}{4}}(z)\right]\right) \ .
\end{equation}
The series expansion provided here can be used to perform analytical studies of the geometry near $z=1$. For that purpose, the computer algebra Mathematica package xAct \cite{JMMG} will be very useful.

\section{Internal geometry and horizons} \label{section:V}

To study the internal geometry of black holes,  it is convenient to use coordinates in which the metric is
well-defined even at the event horizons. For line elements of the form we are using,
$ds^2=-B(r)dt^2+C(r)dr^2+r^2d\Omega^2$, it is particularly useful to replace the $(t,r)$ coordinates by the
so-called Eddington-Finkelstein coordinates $(v,r^*)$ \cite{NS-F} that turn the line element into
\begin{equation}\label{eq:EF}
ds^2=-B(r)dv^2+ 2dv dr^*+r^2(r^*)d\Omega^2 \ ,
\end{equation}
where $v=t+ x$ with $(dx/dr)^2=C(r)/B(r)$, and $r=r(r^*)$ is such that $(dr^*/dr)^2={B(r)C(r)}=1/\sigma_-$.   From our definitions in (\ref{eq:g}), we have $B(r)= \frac{A(z)}{\sigma_+}$, $C(r)=\frac{\sigma_+}{\sigma_- }\frac{1}{A(z)}$, $r=r_c z$, where $r_c=\sqrt{r_q l_P}$, and we can also define $r^*=r_c z^*$ (for notational convenience, from now on we use the dimensionless variables $z$ and $z^*$ instead of $r$ and $r^*$).

The line element (\ref{eq:EF})  puts forward that the geometry is fully characterized by the functions $B(z)$ and $z(z^*)$. The relation between $z$ and $z^*$ can be found by direct integration and is given by $z^*= {_{2}F}_1\left[-\frac{1}{4},\frac{1}{2},\frac{3}{4},\frac{1}{z^4}\right] z$. For $z\gg 1$, we have $z^*\approx z-1/6z^3+\ldots$, whereas for $z\to 1$ we find $z^*\approx z^*_1 +\sqrt{z-1}+\frac{5}{12} (z-1)^{3/2}+\ldots$, where $z^*_1={\sqrt{\pi } \Gamma\left[\frac{3}{4}\right]}/{\Gamma\left[\frac{1}{4}\right]}\approx 0.59907$.  The relation between $z$ and $z^*$ is monotonic and invertible in the region $z\ge 1$. From (\ref{eq:Gz}) we also see that $B(z)$ is only defined for $z\ge 1$. As we pointed out above, this is a dynamical consequence of the theory, not a coordinate problem. To learn more about this point, we need to study the properties of the geometry as the region $z\to 1$ is approached.

\subsection{Region $z\to 1$}

The general expansion of the metric function $B(z)$ of (\ref{eq:EF})  around $z\approx 1$ leads to
\begin{eqnarray}
B(z)&\approx& - \frac{\left(1+\beta  \delta _1\right)}{4\delta _2}\left(\frac{1}{ \sqrt{z-1}}+\frac{9}{4} \sqrt{z-1}-\ldots\right)+\nonumber \\ &+&\frac{1}{2}\left(1-\frac{\delta _1}{ \delta _2}\right)+\left(1-\frac{2\delta _1}{3\delta _2}\right) (z-1)+\ldots \label{eq:gtt_series}
\end{eqnarray}
This expansion shows that the metric component $g_{vv}$ is in general divergent as $z\to 1$. However, it also points out the existence of a particular  mass-to-charge ratio for which the divergence disappears. In fact, if we take $\delta_1=\delta_1^*\equiv-1/\beta$, we get
\begin{eqnarray}\label{eq:gtt_constrained}
B(z)&\approx &\frac{1}{2}\left(1-\frac{\delta_1^*}{\delta _2 }\right)+\left(1-\frac{2\delta_1^*}{3 \delta _2 }\right)(z-1)- \nonumber \\ &-&\frac{1}{2}\left(1-\frac{8\delta_1^*}{5 \delta _2 }\right)(z-1)^2+\ldots   ,
\end{eqnarray}
where, according to (\ref{eq:d1d2}), $\delta_1/\delta_2=r_q/(2l_P)\gg 1$ for macroscopic black holes.
This result implies that when $\delta_1=\delta_1^*$ the line element and the metric components are finite everywhere. It is worth noting that this feature is quite similar to what is found in some NEDs in GR, for which a particular combination between matter and black hole parameters give rise to metrics which are finite everywhere \cite{NEDs-finite}. In our case, it is the nonlinearity in the gravitational action (\ref{eq:action}), as opposed to the nonlinearity in the matter sector of NEDs, which gives rise to this effect. It should be noted that even though in GR with some NEDs the metric may be finite everywhere, some of the curvature invariants always blow up at the center of the solutions, according to a well known theorem holding for electrostatic fields with Maxwell weak field limit \cite{Bronnikov79}. To see what happens in our theory, we consider next the behaviour of the Ricci scalar, Ricci squared, and Kretschmann scalar of the metric $g_{\mu\nu}$ in the regions $z\to 1$ and also $z\gg 1$.

\subsection{Curvature scalars} \label{sec:V.B}

Using the exact solution for the metric found above, we can compute some relevant curvature invariants to extract useful information about the geometry. The simplest such objects are the Ricci scalar $R(g)=g^{\mu\nu}R_{\mu\nu}(g)$,  the Ricci squared scalar $Q(g)\equiv g^{\mu\nu}g^{\alpha\beta}R_{\mu\alpha}(g)R_{\nu\beta}(g)$, and the Kretschmann scalar $K(g)={R^\alpha}_{\beta\mu\nu} (g) {R_\alpha}^{\beta\mu\nu} (g)$, where ${R^\alpha}_{\beta\mu\nu}(g)$ represents the Riemann tensor of the metric $g_{\alpha\beta}$. In GR, these objects have the following exact values
\begin{equation}
R_{GR}=0 \ , \ Q_{GR}=\frac{r_q^4}{r^8} \ , \ K_{GR}=\frac{12 r_S^2}{r^6}-\frac{24 r_S r_q^2}{r^7}+\frac{14 r_q^4}{r^8} \ .
\end{equation}
Though the Ricci scalar in GR is zero,  because the stress-energy tensor of the electromagnetic field is traceless, the other two scalars are nonzero and, in fact, diverge as $r\to 0$, which signals the existence of a strong singularity at $r=0$.

Since the solution that we found for the metric in our theory is given in terms of infinite series expansions, we find it useful to compute the above scalars in the two natural regimes in which those solutions were found, namely, in the region $z\gg 1$ and in the limit $z\to 1$. When $z\gg 1$, we find the following expansions ($r_c\equiv\sqrt{r_q l_P}$ and $z\equiv r/r_c$)
\begin{eqnarray}\label{eq:Rgr}
R(g)&\approx& -\frac{48 r_c^8}{r^{10}}+O\left(\frac{r_c^9}{r^{11}}\right) \\
Q(g)&\approx& \frac{r_q^4}{r^8}\left(1-\frac{16l_P^2}{r^2}+\ldots\right)  \label{eq:Qgr}\\
K(g)&\approx&K_{GR}+\frac{144 r_S r_c^4}{r^9}+\ldots \label{eq:Kgr}
\end{eqnarray}
It is clear that these results recover $GR$ when $r\gg r_c$ (or, equivalently, $z\gg 1$). However, in the region $z\to 1$ the behaviour of those curvature scalars is completely different from the $z\gg 1$ expansions. When $z\to 1$, we find
\begin{eqnarray}
r_c^2 R(g)&\approx &\left(-4+\frac{16 \delta _1^*}{3 \delta _2}\right)+O\left({{z-1}}\right)+\ldots  \\ &-& \frac{1}{2 \delta _2 }\left(1-\frac{\delta _1^*}{\delta _1}\right)\left[\frac{1}{(z-1)^{3/2}}-O\left(\frac{1}{\sqrt{z-1}}\right)\right]  \nonumber,
\end{eqnarray}

\begin{eqnarray}
r_c^4 Q(g) &\approx &  \left(10+\frac{86 \delta _1^2}{9 \delta _2^2}-\frac{52 \delta _1}{3 \delta _2}\right)+O\left({{z-1}}\right)+\ldots  \\
&+&\left(1-\frac{\delta _1^*}{\delta _1}\right)\left[\frac{6 \delta _2-5\delta _1}{3 \delta _2^2 (z-1)^{3/2}}+O\left(\frac{1}{\sqrt{z-1}}\right)\right] \nonumber \\
&+&\left(1-\frac{\delta _1^*}{\delta _1}\right)^2\left[\frac{1}{8 \delta _2^2 (z-1)^3}-O\left(\frac{1}{({z-1})^2}\right)\right] \nonumber ,
\end{eqnarray}

\begin{eqnarray}
r_c^4K(g)&\approx & \left(16+\frac{88 \delta _1^2}{9 \delta _2^2}-\frac{64 \delta _1}{3 \delta _2}\right)+O\left({{z-1}}\right)+\ldots  \\
&+&\left(1-\frac{\delta _1^*}{\delta _1}\right)\left[\frac{2 \left(2\delta _1-3 \delta _2\right)}{3 \delta _2^2 (z-1)^{3/2}}+O\left(\frac{1}{\sqrt{z-1}}\right)\right]+\nonumber \\ &+&\left(1-\frac{\delta _1^*}{\delta _1}\right){}^2\left[\frac{1}{4 \delta _2^2 (z-1)^3}+O\left(\frac{1}{({z-1})^2}\right)\right] \nonumber.
\end{eqnarray}
From these expansions we see that the curvature scalars diverge at $z=1$ except for those configurations whose charge-to-mass ratio satisfies the condition $\delta_1=\delta_1^*$, since in that case all of them become finite. This is an important result whose physical consequences will be explored in more detail later. For now, it should be noted that the avoidance of the singularity is a local, nonperturbative effect that has no impact on the structure of the black hole at distances $z\gg 1$, which quickly tends to that of GR regardless of the particular value of $\delta_1$, as can be seen from the expansions in (\ref{eq:Rgr}), (\ref{eq:Qgr}), and (\ref{eq:Kgr}).

\subsection{Properties of the hypersurface $z=1$}

Consider the normal to a hypersurface $S(v,r^*,\theta,\phi)=$constant, namely, $N= g^{\mu\nu}\partial_\nu S \partial_\mu$. If $S=r$, then $N^\mu=\frac{dr}{dr^*}( 1, B,0,0)$, $N_\mu= \frac{dr}{dr^*}(0,1,0,0)$, and
$N^\mu N_\mu= \left(\frac{dr}{dr^*}\right)^2B=1/C(r)= \sigma_- B(z)$. This result shows that the vector $N$ is spacelike outside the external
horizon, null at the horizon, and timelike inside the horizon except at $z=1$, where it becomes null again (regardless of the value of $\delta_1$)
due to the presence of $\sigma_-$. This implies that the singularity found for $\delta_1\neq \delta_1^*$ lies on a null hypersurface, which contrasts
with the Schwarzschild (spacelike) and Reissner-Nordstr\"{o}m (timelike) singularities  of GR. In the context of GR, null singularities have been
found in the interior of Reissner-Nordstr\"{o}m black holes perturbed by neutral matter \cite{Poisson90}. Further exploration of the connection between these two results shall be done elsewhere.

To learn more about this null hypersurface, we compute now its surface gravity using the Killing vector $l=\partial_t$. In the coordinates (\ref{eq:EF}), which are regular across the external horizon, the components of this vector  are $l^\mu=(1,0,0,0)$ and $l_\mu=(-B, 1,0,0)$. Since the surface gravity $\kappa$ is defined by  $\nabla_\alpha(l^\mu l_\mu)=-2\kappa l_\alpha$ (evaluated at $z=1$), it follows that $2\kappa= \partial_{r^*} B(r)=\sigma_-^{1/2}B_z/r_c$, which leads to
\begin{equation}
\kappa=\left\{ \begin{array}{lr} \lim_{r\to r_c} \frac{\delta_1^*-\delta_1}{8\delta_1^*\delta_2}\frac{1}{r-r_c }  & \text{ if } \delta_1\neq \delta_1^* \\
0  & \text{ if } \delta_1= \delta_1^* \end{array}\right.
\end{equation}
Note that, strictly speaking, the surface gravity only makes sense when evaluated on a horizon. Since in the cases $\delta_1\neq \delta_1^*$ the null surface $z=1$ is singular,  we believe that the divergence of $\kappa$ in those cases is just a manifestation of the breakdown of the geometric description on that surface. On the contrary, the vanishing of $\kappa$ at the horizon $z=1$ when $\delta_1= \delta_1^*$ puts forward the smoothness of the geometry at that location. We refrain ourselves from interpreting these results in a thermodynamic context because this aspect of Palatini theories of gravity has not been considered in the literature with sufficient detail yet.

\subsection{Horizons}

Horizons are located at the points where the metric function $B(r)$ vanishes. From the definitions given above, this happens when the curves $f_1(z)=1+\delta_1 G(z)$ and $f_2(z)=\delta_2 z \sigma_-^{1/2}$ meet. We have already seen analytically that for large black holes, $r_c/r_S\ll 1$, the external horizon lies almost at the same location as in GR. However, since the internal structure of our black holes is different from that of GR the very  existence of an inner horizon is not guaranteed a priori. Moreover, the location of the external horizon for microscopic black holes may also significantly differ from the prediction of GR. For these reasons, in this section we focus on these points to complete our analysis of the internal structure of these black holes.

We note that due to the character of the solutions, given as infinite power series, the best way to determine whether inner horizons exist or not is by using a graphical representation of the functions $f_1(z)$ and $f_2(z)$. From these representations, see Figs. \ref{fig:1}, \ref{fig:2}, and \ref{fig:3},  one realizes that the first terms of the expansions may provide useful information to understand the main features of the various cases of interest. In this sense, it is worth noting that both  $f_1(z)$ and $f_2(z)$ are monotonic functions whose asymptotes are $f_1(z)\sim 1-\delta_1/z$ and $f_2(z)\sim \delta_2 z$, respectively. This implies that if for some $z_0$ we have $f_2(z_0)<f_1(z_0)$ then at some $z_H>z_0$ we will have $f_2(z_H)=f_1(z_H)$, which implies the existence of an horizon (as can be verified graphically). Following this reasoning, we expand $f_1(z)$ and $f_2(z)$ around $z=1$ and identify  $f_1(z)/f_2(z)\ge 1$ as the condition for the existence of an inner horizon in that region. Using the expansions $f_1(z)\approx 1-\frac{\delta_1}{\delta_1^*}+2\delta_1\sqrt{z-1}$ and $f_2(z)\approx 2\delta_2\sqrt{z-1}$, we find  the following cases:

\begin{itemize}
\item {\bf If $\delta_1=\delta_1^*$ then $\frac{\delta_1}{\delta_2}\ge 1$ .} \\
This relation translates into the condition $r_q\ge 2l_P$. Expressing the charge as $q=N_q e$, where $e$ is the electron charge and $N_q$ the number of charges, we can write $r_q=\sqrt{2\alpha_{em}}N_q l_P$, where $\alpha_{em}$ is the electromagnetic fine structure constant.  With this notation, the above condition becomes $N_q\ge  N_q^c\equiv\sqrt{2/\alpha_{em}}\approx 16.55$.
This result means that in order to have an horizon when $ \delta_1=\delta_1^*$, the number of charges must be at least equal or greater than $N_q^c$. For smaller values of the charge one can verify graphically that there are neither inner nor outer horizons, which represents {\it naked core} solutions. When the charge is greater than $N_q^c$ then we have only an external horizon and no inner horizon. All of these solutions are free of curvature singularities, as results from the analysis of section \ref{sec:V.B}.
\begin{figure}[h]
\includegraphics[width=0.45\textwidth]{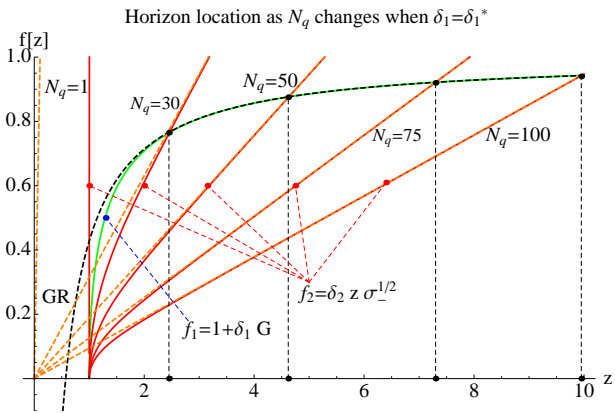}
\caption{The location of the external horizon is given by the
intersection of the curve $f_1$ (solid green) with $f_2$ (solid red
curves labeled by $N_q$). The dashed (orange) straight lines that meet at the origin correspond to $f_2^{GR}=\delta_2 z$. The dashed black curve is $f_1^{GR}=1-\delta_1/z$. Note that $f_1$ and $f_2$ quickly recover the GR behaviour for $z\gg 1$. As a result, the location of the external horizon almost coincides with the GR prediction for $N_q\geq 30$. \label{fig:1}}
\end{figure}

\item {\bf If  $\delta_1<\delta_1^*$ then $\frac{N_q^c}{N_q}\le 1+\frac{\delta_1^*-\delta_1}{2\delta_1^* \delta_1 (z-1)^{1/2}} $ . }\\
This condition indicates that  once the charge-to-mass ratio $ \delta_1<\delta_1^*$ and the number of charges $N_q$ are specified, one  can always find some $z>1$ that verifies the inequality, which implies the existence of an horizon. Therefore, regardless of the value of $N_q$, when $\delta_1<\delta_1^*$ we always have an (external) horizon  (see Fig.\ref{fig:2}).

\begin{figure}[h]
\includegraphics[width=0.45\textwidth]{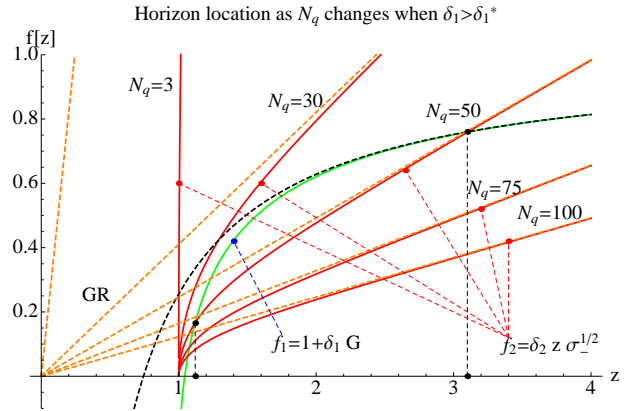}
\caption{Same notation and labeling as in Fig.\ref{fig:1}. When $\delta_1<\delta_1^*$, we have $G(z=1)>0$, which forces all the curves $f_2(z)$ to cut $f_1(z)$ in a single point. Except for very small values of $N_q$, the location of the (external) horizon almost coincides with the GR prediction.  In this plot $\delta_1=\delta_1^*(1-5\times 10^{-1})$.  \label{fig:2}}
\end{figure}

\item {\bf If $\delta_1>\delta_1^*$ then $\frac{N_q^c}{N_q}\le 1-\frac{\delta_1-\delta_1^*}{2\delta_1^* \delta_1 (z-1)^{1/2}}$ . }

From this it follows that for some combinations of $\delta_1$ and $N_q$ there may or may not exist a $z>1$ that satisfies the inequality. This means that in some cases there is no horizon ($f_2(z)>f_1(z)$ always), which implies a naked singularity, while in other cases there may be up to two horizons. This is verified graphically in Fig. \ref{fig:3}, where we can appreciate solutions without horizon, solutions with two horizons, and solutions with one (degenerate) horizon (extreme black hole), a situation analogous to that of the standard Reissner-Nordstr\"om solution of GR.

\begin{figure}[h]
\includegraphics[width=0.45\textwidth]{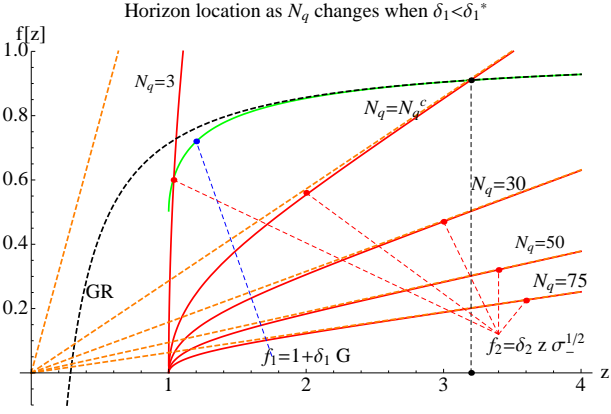}
\caption{Same notation and labeling as in Fig. \ref{fig:1}. When $\delta_1>\delta_1^*$, some configurations have two horizons (see the curves with $N_q=50, 75, 100$) while others have no horizons (like $N_q=30$). The limiting case $N_q=35$ (not shown here) has only one (degenerate) horizon and represents an extreme black hole. The set $N_q<35$ represents naked singularities. In this plot $\delta_1=\delta_1^*(1+3\times 10^{-1})$. \label{fig:3}}
\end{figure}

\end{itemize}

\subsection{Penrose diagrams and analytical extensions}

\begin{figure}[h]
\includegraphics[width=0.45\textwidth]{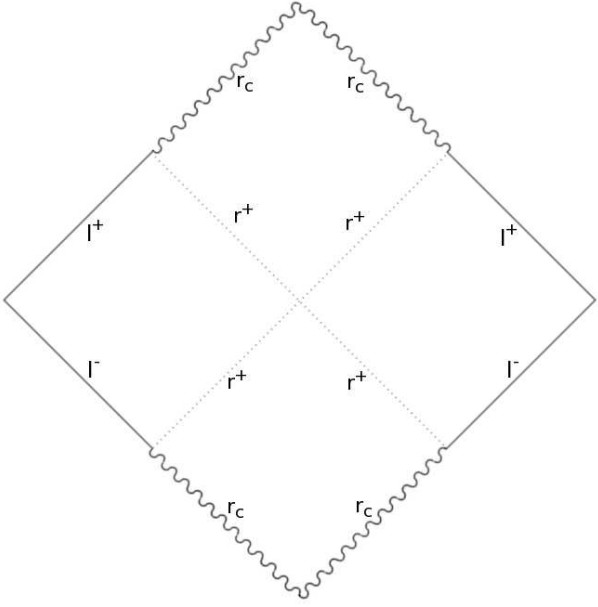}
\caption{Penrose diagram for the case $\delta_1<\delta_1^*$. Unlike in the Schwarzschild black hole, where the $r=0$ singularity  is spacelike, the singularity here is null and appears at $r=r_c$. \label{fig:Penrose_I}}
\end{figure}

The previous discussion about horizons provides useful information for the construction of the Penrose diagrams corresponding to the solutions found.
For instance, we have seen that when $\delta_1<\delta_1^*$, all solutions represent black holes with a single horizon. Behind the horizon we find a  singularity located at $r=r_c$. This configuration is essentially the same as that found in GR for Schwarzschild black holes, except for the fact that the singularity is now null instead of spacelike. As a result, the corresponding Penrose conformal diagram is that represented in Fig. \ref{fig:Penrose_I}.

\begin{figure}[h]
\includegraphics[width=0.45\textwidth]{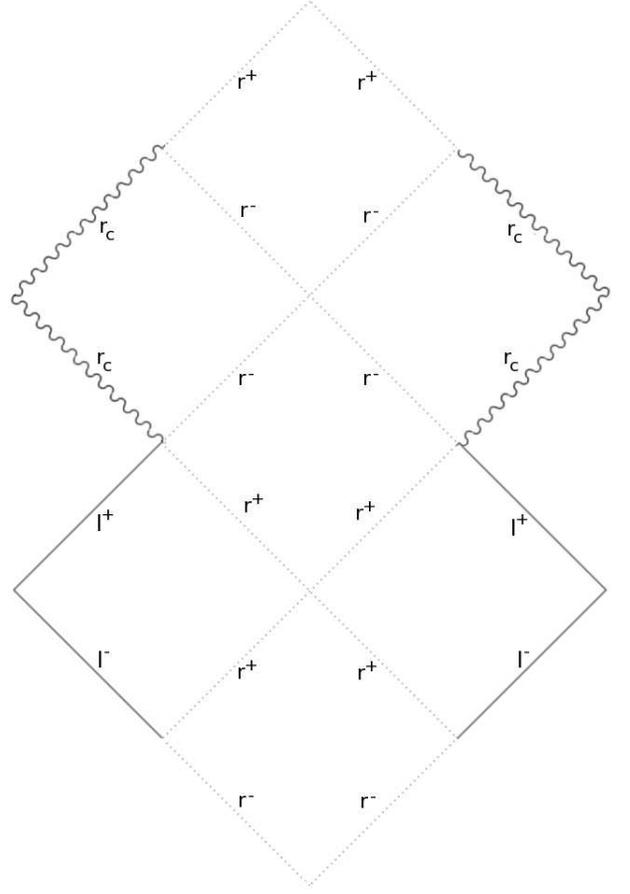}
\caption{Penrose diagram for the case $\delta_1>\delta_1^*$ with two horizons ($r^+$ and $r^-$). The only difference with respect to the GR diagram is that the singularity lies on a null surface at $r=r_c$. \label{fig:Penrose_II}}
\end{figure}

When $\delta_1>\delta_1^*$, we may find the same subcases as in GR, namely, solutions with two horizons, with one double (degenerate) horizon, and naked singularities. Like in the previous example, the main difference is that the singularity is null rather timelike. To illustrate how the Penrose diagram is modified, we plot the case with two horizons in Fig. \ref{fig:Penrose_II}.

\begin{figure}[h]
\includegraphics[width=0.45\textwidth]{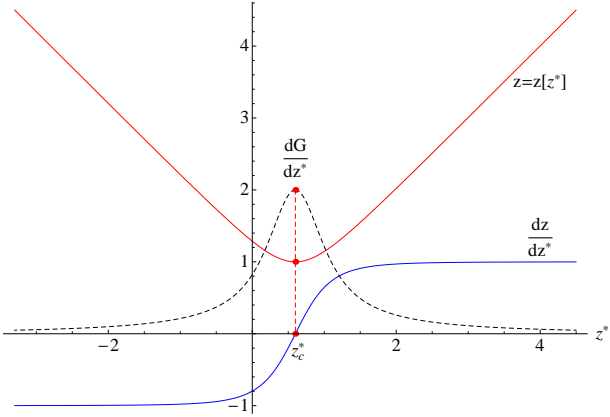}
\caption{Representation of the curves $z=z(z^*)$, $dz/dz^*$ and $dG/dz^*$. The minimum of $z(z^*)$ occurs at $z^*_c\approx 0.599$, where $dz/dz^*$ vanishes and $dG/dz^*$ reaches its maximum value. Recall that $r^*=r_c z^*$. \label{fig:zzstar}}
\end{figure}

\begin{figure}[h]
\includegraphics[width=0.45\textwidth]{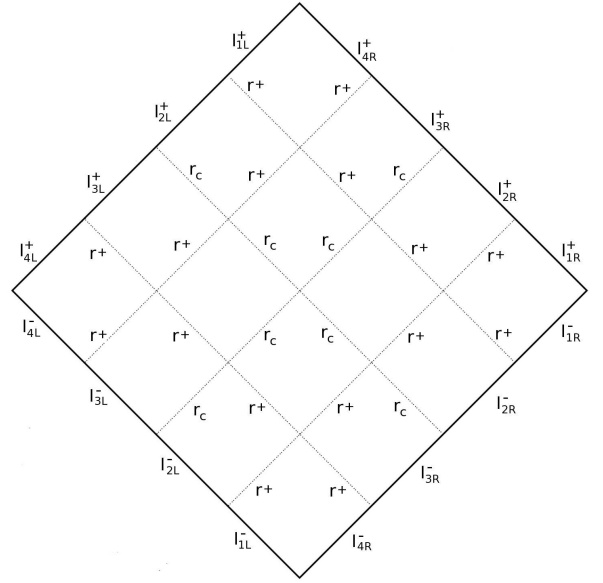}
\caption{Penrose diagram for the nonsingular case $\delta_1=\delta_1^*$. The null surfaces labeled as $r_c$ represent the region where the area of the $2-$spheres reaches its minimum (nonzero) value. An incoming null geodesic (light ray) propagating from $I^-_{1R}$ towards the horizon $r^+$, will reach  $r_c$ and bounce off as an outgoing null geodesic into another region, eventually reaching $I^+_{1L}$ (if no scattering takes place). The surface $r_c$, therefore, plays the role of (attractive) black hole and (repulsive) white hole at the same time. \label{fig:Penrose_III}}
\end{figure}

The case $\delta_1=\delta_1^*$ deserves special attention because the null hypersurface $r=r_c$ is nonsingular. Unlike in the other cases with $\delta_1\neq\delta_1^*$, the absence of a singularity suggests that the geometry  may admit some analytical extension beyond that point. This extension is naturally obtained from the relation $(dr^*/dr)^2=1/\sigma_-$ that defines the function $r^2(r^*)$ in (\ref{eq:EF}). In our analysis following (\ref{eq:EF}), we implicitly assumed that $dr^*/dr=   1/\sigma_-^{1/2}$, and omitted the possibility of having a branch with the negative sign, $dr^*/dr=  - 1/\sigma_-^{1/2}$. In the singular cases $\delta_1\neq\delta_1^*$, the omission of the branch with $dr^*/dr<0$ is justified because there is no natural way to extend the geometry beyond the singularity at $r=r_c$.  However, if there is no singularity at $r=r_c$, the divergence of $dr^*/dr$ at this point simply states that the function $r(r^*)$ has reached a minimum at the point $r^*_c=0.59907 r_c$ (see the discussion following Eq.(\ref{eq:EF}) and  Fig. \ref{fig:zzstar}). For values of $r^*<r^*_c$, the branch with  $dr^*/dr<0$ describes a new region in which the area of the $2-$spheres grows  as $r^*\to -\infty$.  The relation between $r$ and $r^*$ can thus be written explicitly as follows (see Fig. \ref{fig:zzstar})
\begin{equation}
r^*=\left\{\begin{array}{lr}
 {_{2}F}_1\left[-\frac{1}{4},\frac{1}{2},\frac{3}{4},\frac{r_c^4}{r^4}\right] r & \text{ if } r^*\ge r^*_c \\
2r^*_c-{_{2}F}_1\left[-\frac{1}{4},\frac{1}{2},\frac{3}{4},\frac{r_c^4}{r^4}\right] r & \text{ if } r^*\le r^*_c
\end{array}\right.
\end{equation}
In terms of $r^*$, we also have $dG/dz^*= \sigma_+/z^2$, which gives continuity to the metric across the bounce.
Using $r_1$ and $r_2$ to parameterize the $2-$spheres when $r^*>r^*_c$ and $r^*<r_c^*$, respectively, the line element (\ref{eq:EF}) can be written as
\begin{equation}\label{eq:EF2}
ds^2=\left\{\begin{array}{lr}
        -B(r_1)dv^2+ \frac{2}{\sigma_-^{1/2}}dv dr_1+r_1^2d\Omega^2 & \text{ if } r^*>r^*_c \\
-B(r_2)dv^2- \frac{2}{\sigma_-^{1/2}}dv dr_2+r_2^2d\Omega^2 & \text{ if } r^*<r^*_c \end{array}\right.
\end{equation}
This representation is useful to understand that for future directed ($dv>0$) worldlines, $drdv\leq 0$ on timelike or null worldlines if $r^*>r^*_c$, but $drdv\geq 0$ if $r^*<r^*_c$.  From a physical point of view, this means that a spherical shell of matter that collapses and crosses the external horizon will unavoidably shrink to a sphere of area $4\pi r_c^2$ before bouncing off into an outgoing shell of increasing area. The outgoing shell, obviously, cannot return to the region from which it was sent initially because it crossed an event horizon. As can be seen from the Penrose diagram of this spacetime in Fig. \ref{fig:Penrose_III}, the outgoing shell can reach several different final regions (different $I^+$ regions for light rays).

\section{Physical aspects} \label{section:VI}

We now discuss several points useful to understand some physical aspects of the solutions described in previous sections.

\subsection{Quantum nature of the nonsingular solutions}

We first note that using $r_q=\sqrt{2\alpha_{em}} N_ql_P$ we find that the area of the $r=r_c$ surface is given by $A_{core}= N_q\sqrt{2\alpha_{em}} A_P$,  where $A_P=4\pi l_P^2$ is Planck's area. This admits a nice physical interpretation since it suggests that each charge sourcing the electric field has associated an elementary quantum of area of magnitude $\sqrt{2\alpha_{em}} A_P$. From this it follows that the ratio of the total charge $q$ by the area of this surface gives a universal constant, $\rho_q=q/(4\pi r_{c}^2)=(4\pi\sqrt{2})^{-1}\sqrt{c^7/(\hbar G^2)}$, which up to a factor $\sqrt{2}$ coincides with the Planck surface charge density. This result is independent of the mass of the black hole and, therefore, applies both to singular as well as to regular solutions.

If we focus now on the regular solutions only, we find that the regularity condition $\delta_1=\delta_1^*$ sets the following mass-to-charge relation
\begin{equation}\label{eq:rq-rs}
r_S=\frac{1}{2\delta_1^*}\sqrt{\frac{r_q^3}{l_P}} \ \leftrightarrow  \  \frac{M_0}{(r_ql_P)^{3/2}}=\frac{1}{4\delta_1^*}\frac{m_P}{l_P^3}\ ,
\end{equation}
which can be interpreted in the sense that the matter density inside a sphere of radius $r_{c}=(r_ql_P)^{1/2}$ becomes another universal constant

\begin{equation}
\rho_{core}^*=\frac{M_0}{V_{core}}=\frac{\rho_P}{4\delta_1^*}.
\end{equation}
From the definition of $\delta_1$ and $\rho_{core}$, comparing any two configurations one can verify that
\begin{equation}
\frac{\delta_1^{(a)}}{\delta_1^{(b)}}=\frac{\rho_{core}^{(b)}}{\rho_{core}^{(a)}} \ .
\end{equation}
This relation is very useful to classify the different black hole configurations, because the relation $(1+\beta \delta_1)$ that appears in
(\ref{eq:gtt_series}) can be written as $(1-\delta_1/\delta_1^*)=(1-\rho_{core}^*/\rho_{core})$,  where $\rho_{core}^*$ is the density of
the regular core. This representation allows for a more  physical interpretation and classification of  the solutions given in Sec. \ref{section:V}.
In particular, the nonsingular configuration $\rho_{core}=\rho_{core}^*$ is the only case in which the core density
is a magnitude independent of $M_0$ and $q$ and, in fact, turns out to be given in terms of fundamental constants
only. This fact puts forward the special quantum-gravitational nature of the nonsingular solutions. In this sense,
we believe that the regularity condition  $\rho_{core}=\rho_{core}^*$, rather than as a fine-tuning issue, should
be interpreted in the spirit of a quantum constraint relating the mass and charge (or charges, in general) of the
solutions to avoid the singularity. It should also be noted that adiabatic changes of the charge and mass of the black hole \cite{Bekenstein97} do not allow transitions between  nonsingular configurations, which indicates that such configurations have different quantum numbers (different $q$ and $M_0$ but the same $\rho_{core}^*$). This must have important consequences for Hawking radiation, because if the emission of quanta is to be compatible with the regularity of the core, then the spectrum must necessarily have a discrete structure. We shall leave the exploration of this issue for future works.

\subsection{Physical and analytical extensions of the electrostatic solution}

The above results picture a black hole interior with an ultracompact core of radius $r_{c}$ which contains all the mass in its interior and all the charge on its surface. This view is physically very appealing but should be compatible with the mathematical solution represented in Fig. \ref{fig:Penrose_III}. In fact, Fig. \ref{fig:Penrose_III} represents an exact mathematical solution of a physically incomplete problem, because it does not take into account the necessary existence of the massive charged particles that generate the electrostatic field. The fact that the surface $r=r_c$ is null suggests that if massive charged particles were explicitly included in the problem, then the Killing vector field $\partial_t$ could become again timelike in the region hidden by the $r=r_c$ horizon, where the sources should be located, thus allowing for the existence of static interior solutions of the type suggested by the constraint (\ref{eq:rq-rs}).  Therefore, for the description of the geometry behind the $r=r_c$ horizon, one should  specify the ${T_\mu}^\nu$ of the sources that carry the mass of the core and the charge that generates the external electric field, which would allow to define a new auxiliary metric $\tilde{h}_{\mu\nu}$ able to parameterize the internal geometry of the core (assuming that suitable matching conditions can be found at $r=r_c$).  Since the regularity condition (\ref{eq:rq-rs})  supports that the core matter density is bounded, we expect the existence of completely regular solutions in agreement with the results found for this same theory of gravity in cosmological scenarios \cite{Barragan2010}.

It is worth pointing out that if the $r=r_c$ null surface could be smoothly matched to an interior region filled with matter, where $\partial_t$ were timelike, then the analytical extension of the exterior spacetime beyond $r=r_c$ would admit two possible and different branches, which would appear on different sheets of a larger conformal diagram. One sheet would contain the matter-filled region and another the analytical extension shown in Fig. \ref{fig:Penrose_III}, where the radial coordinate {\it bounces}.  This could give complete physical reality to the spacetime of Fig. \ref{fig:Penrose_III} in such a way that particles approaching $r=r_c$ would be scattered into the {\it white hole} region instead of falling into the matter-filled sector. This and related phenomenological issues will be explored elsewhere.

\subsection{Astrophysical black holes}

From the large $z$ expansion in (\ref{eq:gtt-far}) and (\ref{eq:grr-far}), we saw that the GR solution $B^{GR}(r)=1-\frac{r_S}{r}+\frac{r_q^2}{2r^2}$ is an excellent approximation for any $r\gg l_P$. This implies that the location of the external horizon of these charged black holes is essentially the same as in GR. We thus find
\begin{eqnarray}  r_+&=&\frac{r_S}{2}\left(1+\sqrt{1-2r_q^2/r_S^2}\right)= \\
&=&\frac{r_S}{2}\left(1+\sqrt{1-4\delta_1^*/(N_q\sqrt{2\alpha_{em}})}\right) \nonumber,
\end{eqnarray}
where (\ref{eq:rq-rs}) has been used. For a solar mass black hole, where the number of protons is around $N_{p,\odot}\sim 10^{57}$, Eq.(\ref{eq:rq-rs}) implies that the number of charges needed to avoid the $z=1$ singularity  is just $N_{q,\odot}=(2r_S \delta_1^* /l_P)^{2/3}/\sqrt{2\alpha_{em}}\approx 2.91 \times 10^{26}$ (or $\sim 484$ moles), which is a very tiny amount on astrophysical terms. In fact, this amount of charge is comparatively so small, $N_{q,\odot}/N_{p,\odot}\sim 10^{-31}$, that it seems reasonable to expect that a quantum gravitational violation of electric charge conservation could naturally act to avoid black hole singularities in stellar collapse processes. In general, $N_q=N_{q,\odot} (M/M_\odot)^{2/3}$ implies that in astrophysical scenarios $r_+\approx r_S$ (see Figs. \ref{fig:1}, \ref{fig:2} and \ref{fig:3} ).

\subsection{Microscopic black holes}

\begin{figure}[h]
\includegraphics[width=0.45\textwidth]{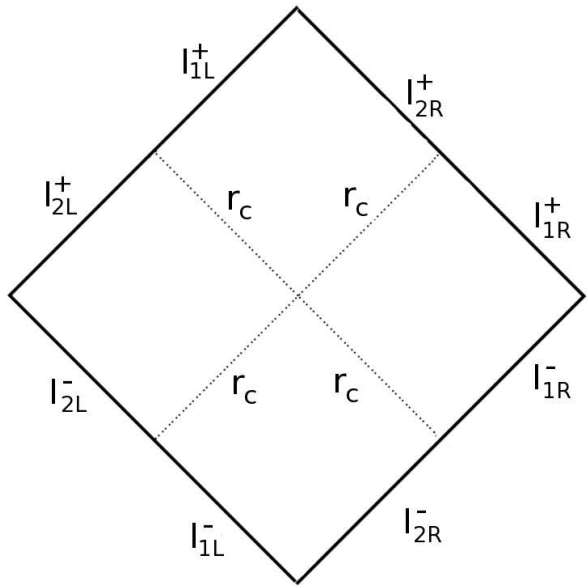}
\caption{Penrose diagram for the nonsingular case $\delta_1=\delta_1^*$ without event horizon, $N_q<N_q^c$.
If the asymptotic region $I^+_{1L}$ is identified with $I^+_{1R}$, then the surface $r_c$ should behave as a perfectly reflecting boundary. In the general case, however, light rays reaching $r_c$ would go through into a new universe.
\label{fig:Penrose_IV}}
\end{figure}

We have already seen that as $N_q$ drops below the critical value $N_q^c=\sqrt{2/\alpha_{em}}\approx 16.55$, the external horizon disappears and the core undresses becoming directly {\it observable} (see Fig. \ref{fig:1}). The Penrose diagram corresponding to the maximal analytical extension of these naked regular cores, $N_q<N_q^c$, is depicted in Fig. \ref{fig:Penrose_IV}.

From (\ref{eq:rq-rs}), it follows that the mass of these objects is
\begin{equation}\label{eq:spectrum}
M=\frac{N_q^{3/2} (2\alpha_{em})^{3/4} m_P}{4\delta_1^*}\approx \frac{N_q^{3/2}m_P}{55}.
\end{equation}
For the particular case $N_q=N_q^c$, we find $M^c=m_P/(\sqrt{2}\delta_1^*)=m_P \pi^{3/2}/(3\Gamma[3/4]^2)\approx 1.23605 m_P$.
It is worth mentioning that this very particular number also appears in the computation of the total energy associated to the electric field in Born-Infeld nonlinear electrodynamics, which indeed is found to be $\varepsilon(q)=\pi^{3/2}/(3\Gamma[3/4]^2) \beta^{1/2}q^{1/2}$ with $\beta$ the Born-Infeld parameter \cite{NEDs-finite}. Its presence here puts forward the regularizing role played by gravitation. Moreover, as the electromagnetic field itself is expected to receive corrections at high energies (see \cite{string-NED} for some string theory results in  this regard), it also points out that an improved description of the electromagnetic field through nonlinear electrodynamics in these scenarios could provide interesting new insights on the interaction between the nonlinearities of matter and gravitation. All these facts suggest that the theory (\ref{eq:action}) may shed new light on the problem of sources in electrodynamics coupled to gravitation \cite{OR2011b,OR2012b}, which already arises in the Reissner-Nordstr\"om and Schwarzschild solutions of GR \cite{Ortin}.

Finally it is worth noting that if $N_q$ is seen as an integer number, besides the core area, the mass of these nonsingular naked cores and black holes is quantized, as has been recently claimed in \cite{Dvali10a} on very general grounds. As pointed out before, one thus expects a discrete spectrum of Hawking radiation, because physically allowed transitions should occur only between regular configurations. This illustrates how Planck-scale physics may affect the perturbative predictions of the semiclassical approach. New investigations on all these issues are currently underway and shall be published elsewhere.

\subsection{Our model as a deformation of GR}

From the analysis of the geometry in the $z\to 1$ region, we have found that the solutions of our theory can be classified into three types according to the value of $\delta_1$ or, equivalently, according to the value of the core density $\rho_{core}$. When $\rho_{core}>\rho_{core}^*$, then the conformal diagram of Fig. \ref{fig:Penrose_I} shows a structure very similar to that found for the Schwarzschild black holes of GR. This admits a nice physical interpretation because when the core contains more mass than in the nonsingular case, the black hole looks more like an uncharged object, such as a Schwarzschild black hole. On the other hand, when the mass-to-charge ratio  is smaller than the expected value of a regular configuration, $\rho_{core}<\rho_{core}^*$, then the resulting structure is closer to a typical Reissner-Nordstr\"{o}m black hole of GR (see Fig. \ref{fig:Penrose_II}). Our black holes, therefore, are somehow smoothly interpolating between the abrupt structures found in GR. The nonsingular solutions, $\rho_{core}=\rho_{core}^*$, lie in the middle of these two disconnected branches (Schwarzschild - Vs - Reissner-Nordstr\"{o}m), and represent a kind of object not found in GR but that emerges when Planck scale effects are incorporated in the problem.

The smooth interpolation found here between the Schwarzschild and Reissner-Nordstr\"{o}m solutions of GR shares some resemblance with the behaviour of the contracting and expanding branches of the nonsingular cosmological solutions found in this Palatini $f(R,Q)$ model \cite{Barragan2010}. In GR, one can classify cosmological solutions in two groups, one corresponding to an expanding branch that starts in a (big bang) singularity, and its time reversal, which corresponds to a contracting branch that ends in that (big crunch) singularity. In the quadratic Palatini model studied here, the two singular branches of GR are smoothly connected through a bouncing solution that avoids the singularity.

\section{Summary and conclusions} \label{section:VII}

In this work we have studied the structure of electrically charged black holes in a Palatini extension of GR characterized by a Ricci-squared term. Theories of this type could be naturally motivated by quantum effects in curved spacetimes. These theories provide modified dynamics without introducing new dynamical degrees of freedom. This implies, in particular, that the resulting solutions can be completely classified using the same parameters as one finds in GR, namely, the total charge $q$ and total mass $M_0$. The absence of new dynamical degrees of freedom also guarantees that these theories are free from ghosts and other potential dynamical instabilities.

We have obtained exact analytical solutions expressed as infinite power series expansions. These solutions show that the structure of these black holes coincides with that of the well known Reissner-Nordstr\"{o}m black holes of GR for values of $r\gg r_c$, where $r_c=l_P \sqrt{2N_q/N_q^c}$, with $N_q$ representing the number of charges and $N_q^c\approx 16.55$.  Important modifications arise as we approach the minimum of the radial coordinate, the region $r\to r_c$.
At this radius, the gauge invariant quantity $X\equiv -\frac{1}{2}F_{\mu\nu}F^{\mu\nu}$
that represents the energy density of the electromagnetic field reaches its maximum value $X_{max}= \rho_P c^2/2$, and other  quantities of interest such as ${T_\mu}^\nu$ also take (finite) Planck scale values. This means that the modified dynamics of our Palatini $f(R,Q)$ model has the effect of setting upper bounds at the Planck scale on the energy density of the matter fields involved, a property already observed in cosmological settings.

We have found that $r=r_c$ is a singular null hypersurface if $\delta_1\neq \delta_1^*$ (see  (\ref{eq:d1d2}) for the definition of $\delta_1$). When the mass-to-charge ratio $\delta_1$ is set to the particular value $\delta_1^*=3\Gamma[3/4]^2/\sqrt{2\pi^3}\approx  0.572$, then this null surface becomes nonsingular and the geometry can be analytically extended by means of a {\it bounce} of the radial coordinate (see Fig. \ref{fig:zzstar}).
We pointed out that the area of the null hypersurface $r=r_c$ grows linearly with the number of charges, $A_{core}= N_q\sqrt{2\alpha_{em}} A_P$. As already mentioned, this behaviour suggests that each charge sourcing the electric field has associated an elementary quantum of area of magnitude $\sqrt{2\alpha_{em}} A_P$. Using this result, direct computation of the surface charge density,  $\rho_q=q/(4\pi r_{c}^2)$, gives $\rho_q=(4\pi\sqrt{2})^{-1}\sqrt{c^7/(\hbar G^2)}$, which up to a factor $\sqrt{2}$ coincides with the Planck surface charge density. For the nonsingular solution, the condition $\delta_1= \delta_1^*$ can be seen as indicating that the  mass density of the core is $\rho_{core}^*=\rho_P/4\delta_1^*$, i.e., it is of order the Planck mass density. It must be noted that $\rho_q$ and $\rho_{core}^*$ are given in terms of the  fundamental constants $\hbar, G,$ and the speed of light $c$, and are independent of $q$ and $M_0$ (in the singular cases, $\rho_{core}$ does depend on $q$ and $M_0$). In our opinion, this is a clear manifestation of the quantum gravitational nature of the nonsingular solutions. Rather than as a fine-tuning problem, this constraint on the core density should be seen as a quantization condition that selects a discrete set among all the classically allowed solutions. In a sense, this is analogous to Bohr's atomic model, where the stability of Hydrogen under electromagnetic emission of radiation was postulated assuming the existence of certain privileged orbits that had to satisfy specific quantization conditions.

On the other hand, we have found the mass spectrum given in (\ref{eq:spectrum}), which is valid for all (positive) values of $N_q$. This mass spectrum has important implications for the emission of Hawking quanta. This is so because if physically allowed transitions occur between nonsingular configurations only, then the resulting spectrum must be discrete. Our analysis also puts forward the existence of a new kind of nonsingular objects which are not hidden by an external horizon. These {\it naked cores} exist for values of the charge comprised within the interval $0<N_q<N_q^c$.

An important lesson that follows from our analysis is that the boundedness of the energy density does not necessarily imply that the spacetime is nonsingular. In order to find nonsingular solutions, it is necessary that charge and mass satisfy a particular relation. In this sense, it seems fair to say  that the mass spectrum and sizes of the nonsingular objects described here need not be in correspondence with actual physical particles.  From the discussion of section \ref{section:IV} on the relation between $r^2$ and $\tilde{r}^2$, it follows that including new scales in the problem (such as the masses and different gauge charges of the particles making up the system) should have an impact on the resulting value of $r_c$, defined as the value of $r$ at which the energy density reaches its maximum. The analogous of the regularity condition $\delta_1=\delta_1^*$ could also set more complicated constraints between the total mass and total charges of the system, thus providing a richer structure to the set of nonsingular solutions.  These aspects, together with the process of formation and the stability under perturbations of the nonsingular solutions (with and without external horizon) studied here is currently underway.

\acknowledgments

The work of G. J. O. has been supported by the Spanish grant FIS2008-06078-C03-02, FIS2011-29813-C02-02, the Consolider Program CPAN (CSD2007-00042), and the JAE-doc program of the Spanish Research Council (CSIC). D. R. -G. thanks the hospitality of the theoretical physics group at Valencia U., where part of this work was carried out. We are very grateful to  A. Fabbri and J. Navarro-Salas for their useful comments and illuminating discussions.

\end{document}